\pdfoutput=1
\documentclass[review]{elsarticle}
\usepackage{geometry}
\usepackage{lineno,hyperref}
\usepackage{amsfonts}
\usepackage{amsmath}
\biboptions{numbers,sort&compress}
\modulolinenumbers[5]

\journal{Journal of \LaTeX\ Templates}

\begin{document}

\begin{frontmatter}

\title{Modeling the time-periodicity of in-degree distributions in scientific citation networks}



\author[mymainaddress]{Qi Liu\corref{mycorrespondingauthor}}
\cortext[mycorrespondingauthor]{Corresponding author}
\ead{liuqi@smail.nju.edu.cn}

\author[mymainaddress]{Zheng Xie}
\author[mymainaddress]{Zonglin Xie}
\author[mymainaddress]{Engming Dong}
\author[mymainaddress]{Jianping Li}

\address[mymainaddress]{College of Science, National University of Defense Technology, Changsha, Hunan, China}

\begin{abstract}
In a range of citation networks, 
the in-degree distributions boast time-periodicity---the  distributions of citations per article published each year present similar scale-free tails. This phenomenon can be regarded as a consequence of the emergence of hot topics and the existence of the ``burst'' phenomenon.
   With this inference considered, a geometric model based on our previous study is established, in which the sizes of the influence zones of nodes follow the same power-law distribution and decrease with their ages.
 The model successfully reproduces the time-periodicity of the in-degree distributions of the empirical data, and accounts for the presence of citation burst as well. Moreover, a reasonable explanation for the emergence of the scale-free tails by regarding the citation behavior between articles as a ``yes/no'' experiment is presented. The model can also predict the time-periodicity of  the local clustering coefficients, which indicates that the model is a good tool in researches on the evolutionary mechanism of citation networks.
\end{abstract}


\end{frontmatter}

\section*{Introduction}
Articles and their citations can be pictured as a network\cite{liu5,liu6}, in which articles are regarded as nodes and a directed edge will be drawn from article $i$ to article $j$ if $i$ cites $j$. Those graphs are acyclic. Since new articles usually only cite the published articles\cite{lq2}, directed loops cannot be obtained in this way.

A growing body  of literature on citation networks has focused on the research of the probability distribution of the number of citations per article, i.e. the in-degree distribution.
In his articles, Price\cite{liu5,liu6} pointed out  the cumulative advantage  to explain the scale-free tails of the in-degree distributions of citation networks. He believes that the rate at which an article is cited  should be proportional to the citations that it already has.
George et al\cite{lq1} did not only penetrate into  the scale-free tails of the in-degree distributions of citation networks but also the forepart behavior and the tipping point between the two different regions via a two-mechanism model.
Moreover, Xie et al\cite{lq4,lq5,lq9,liu14} contributed to the study of in-degree distributions of citation networks by proposing a series of geometric models.
The aforementioned studies  all serve to  reveal the fundamental  feature of the in-degree distributions that the majority of articles are rarely cited while few  articles are frequently cited. However,  some problems remained unanswered: What is the publication time of the highly cited articles? Are there such articles published each year? 

\begin{table}[!t]
\renewcommand{\arraystretch}{1.3}
\setlength{\abovecaptionskip}{0pt}
\setlength{\belowcaptionskip}{10pt}
\caption{\textbf{\small The typical statistic features of the analyzed networks.} In the header of the table, ASP, CC, PG and MO denote the average shortest path (if the network is not connected, the giant component can be  calculated), the clustering coefficient, the node proportion of giant
component and modularity, respectively. The first two networks are put forward in the articles (which are published from 2002-01 to 2015-12) of Science and Nature. The third network stems from Proceeding of the National Academy of Science of the United States of America (PNAS) covering articles from 2000-01 to 2015-11. The last network is generated by our model, the parameters of which are $m=12$, $T=100$, $\beta=0.88$, $\alpha=0.6$, $\gamma=0.9$, $\eta=300$, $p=1$. }
\centering

\begin{tabular}{l c c c c c c c } \hline\hline
Networks&Nodes& Links&ASP&CC &PG &MO \\ \hline
Science &36642 &28460 &11.3270 &0.1374  &0.3855 &0.9178\\
Nature   &36787 &34949 &10.6000   &0.1255  &0.4198 &0.8916\\
PNAS  &55358 &86534 &9.9328 &0.1178    &0.7975&0.8706  \\
Modeled network &60600 &87053& 14.5063&0.2064&0.0701&0.9927  \\
\hline\hline
  \end{tabular}
  \label{table1}
\end{table}
In fact, highly cited articles emerge each year. A more interesting phenomenon is that the in-degree distributions of citation networks boast time-periodicity---the in-degree distributions of articles published each year in many citation networks boast similar scale-free tails. The overall in-degree distributions of citation networks have been successfully predicted in our previous study\cite{lq9} in which the citation attractiveness of an article is expressed as a geometric area, whose size is inversely proportional to its birth time. To penetrate into the evolutionary mechanism of citation networks leading to this phenomenon,
 a new concentric circles model is developed based on the previous model.
 In the new model, the sizes of the geometric areas of nodes (articles) follow the same power-law distribution. Considering the fact that novelty of an article will  eventually wear off, these sizes decrease with the age of nodes.
The model serves to reproduce the features of in-degree distributions of articles published each year of the empirical data and accounts for the presence of the citation burst as well.
Moreover, in a reasonable interpretation for the emergence of the similar scale-free tails of those distributions,  whether one article cites another or not is regarded  as a ``yes/no'' experiment\cite{liu14}.
Besides, we also examine the ability of  the model to predict the relation between the local clustering coefficients of articles published each year and their in-degrees.

This article is structured as follows. Some real data are introduced in Section 2.  The model is illustrated in Section 3. The time-periodicity of the in-degree distribution and that of the local clustering coefficients are analysed in Section 4 and Section 5. The conclusion is given in the last section.
\begin{figure}
\setlength{\abovecaptionskip}{0.cm}
\setlength{\belowcaptionskip}{-0.cm}
\centering
 \includegraphics[height=1.5in,width=6in,angle=0]{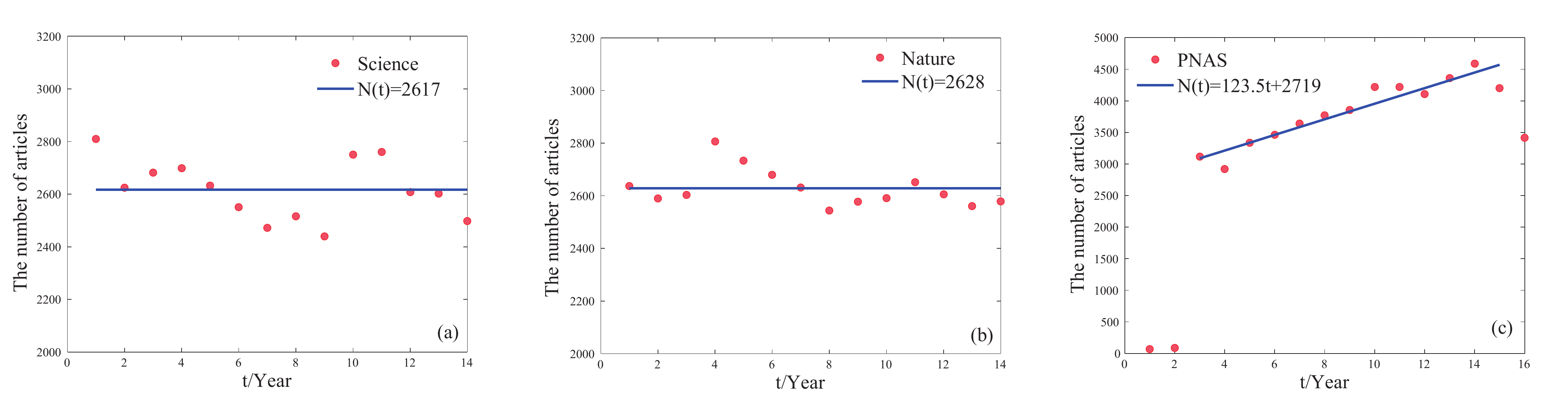}
\caption{\small\textbf{The changing trend of the  number of articles published each year in Table~\ref{table1}.} Panels (a,b) exhibit the trends of articles published in Science and Nature during 2002-2015, which can be approximately fitted by constant functions. Panel (c) shows the trend of articles published in PNAS during 2000-2015, which can be approximately fitted by a linear function. }
\label{fig1}       
\end{figure}
\section*{Data}
Three empirical networks (Table~\ref{table1}) mined from web of science (http://www.webofscience.com) are analysed in this paper. One of the networks is from Science, including  36642 articles published during 2002-2015. Another network is from Nature, constituting 36787 articles published during 2002-2015. The evolutionary trends of the number of articles published each year in these two citation networks are approximately constant (Figure~\ref{fig1}a,\ref{fig1}b). The last citation network derived from Proceeding of the National Academy  of Science of the United States of America (PNAS) covers 55358 articles from January 2000 to November 2015.  An increase is palpable in the changing trend of the number of articles published each year, which  increases in an approximately linear fashion (Figure~\ref{fig1}c),  despite the numbers of articles published in 2000, 2001 and 2015 deviate from the fitting curve due to the absence of articles in the data source.


Since some properties of the articles published later in the empirical data have not evolved completely, the study of the properties of these articles is bypassed. And those articles with palpable properties are highlighted.

\begin{figure}
\setlength{\abovecaptionskip}{0.cm}
\setlength{\belowcaptionskip}{-0.cm}
\centering
 \includegraphics[height=3.2in,width=5.5in,angle=0]{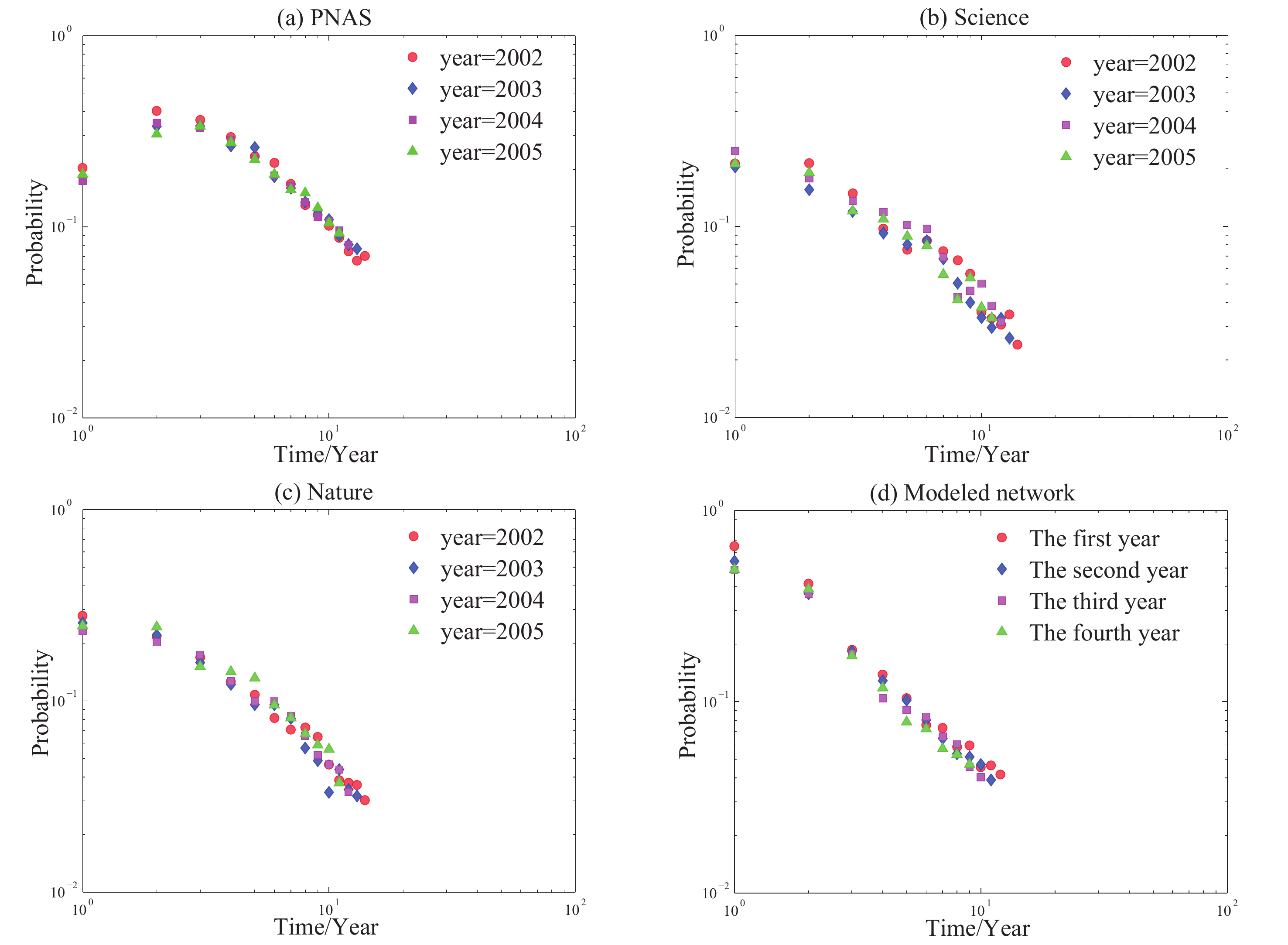}
\caption{\small\textbf{The probabilities that articles published in one year receive citations from articles published in another year.}  Panels (a,b,c) show the relations between those probabilities of articles published in 2002, 2003, 2004 and 2005 and their ``ages'' according to the empirical data. Panel (d) shows the relation between those probabilities of articles published in the first four years and their ``ages'' in the modeled network in Table~\ref{table1}.  }
\label{fig_burst}       
\end{figure}
The probability that articles published in one year receive citations from ones published in another year is calculated to evaluate the rate of citations that articles published in one year can  get. It shows that articles in these three citation networks are highly likely to have received citations within a  few years since their publication while  beyond this period the likelihood  decreases (Figure~\ref{fig_burst}a,\ref{fig_burst}b,\ref{fig_burst}c), which demonstrates the existence of the ``burst'' phenomenon.
Meanwhile, a comparison of the in-degree distributions of articles published in 2002, 2004, 2006 and 2008 has been drawn. Surprisingly, it (Figure~\ref{fig2}a,\ref{fig2}b,\ref{fig2}c) turns out that  the in-degree distributions in the empirical data are  periodic in the time dimension---the in-degree distributions of articles published each year boast the similar scale-free tails. It means that there are highly cited articles each year. The local clustering coefficients of articles  in these empirical data also boast the time-periodic property (Figure~\ref{fig5}a,\ref{fig5}b,\ref{fig5}c)---the large in-degree articles published each year have low local clustering coefficients while they are higher for small ones.

\section*{Model}
The scale-free property of the in-degree distributions of citation networks is usually interpreted as a consequence of the preferential attachment in substantial previous studies. Nevertheless, in those models only based on the preferential attachment (or cumulative advantage), e.g. the Price model, the nodes with large in-degree are always born early\cite{lq10}, which runs counter to the case that there exist highly cited articles each year. The same problem also exists in our previous study\cite{lq9}.
As a result, it is necessary to identify a new mechanism to illuminate why the in-degree distributions of articles published each year in the empirical data have similar scale-free tails.
Acoordingly, based on two points of cognition, a promotion of the model we have built\cite{lq9} is completed:
\begin{figure}
\setlength{\abovecaptionskip}{0.cm}
\setlength{\belowcaptionskip}{-0.cm}
\centering
 \includegraphics[height=2.1in,width=3in,angle=0]{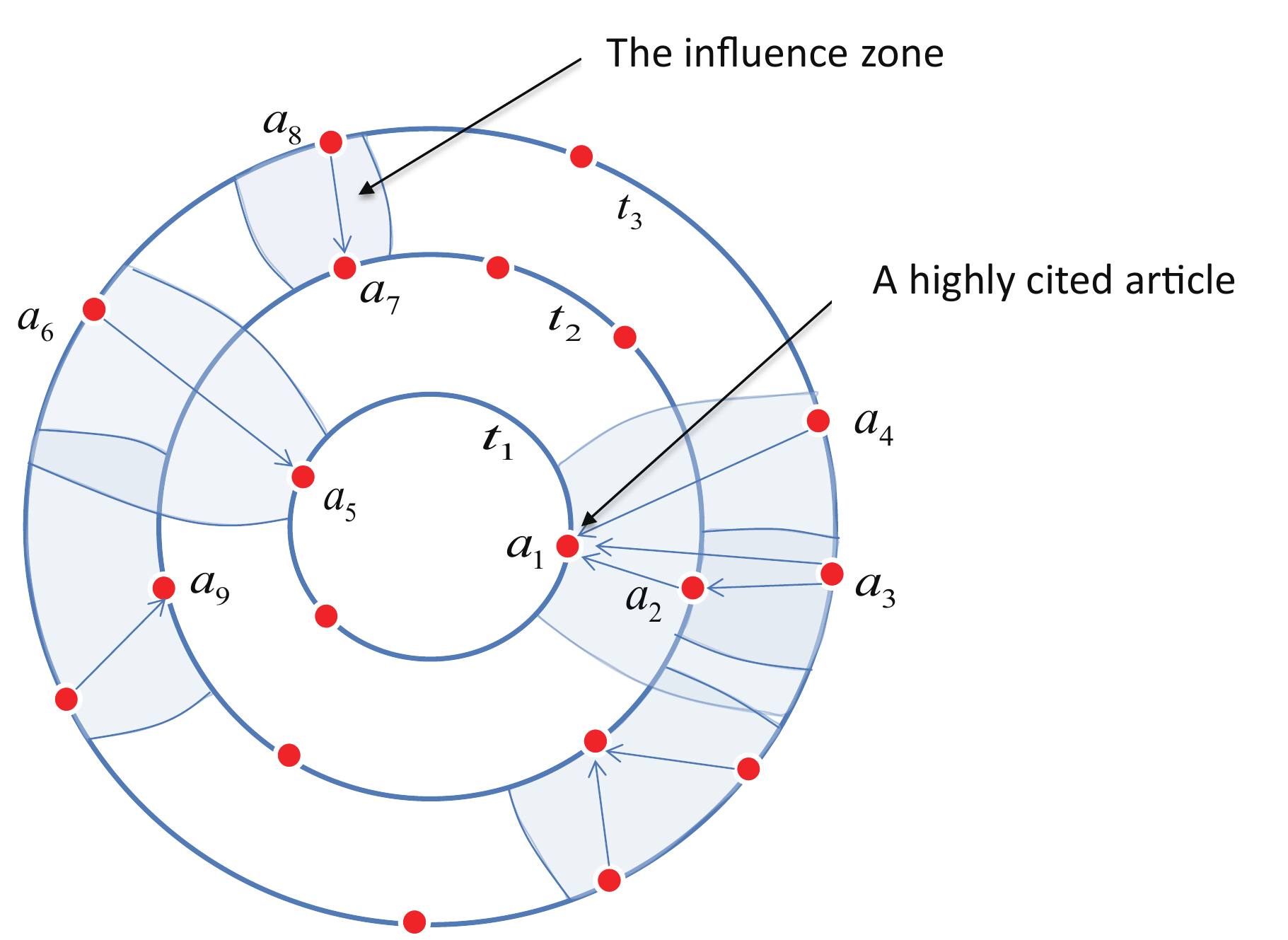}
\caption{\small \textbf{Diagram of the model}. The red dots on each circle denote the articles published in the same issue, the number of which increases linearly. The blue areas represent the influence zones of nodes, which decay with time. Diversities of  the academic influences are expressed by different zonal sizes following the same distribution, e.g. the size of the influence zone of article $a_{9}$ is bigger than that of other's. The edges in the model are linked according to the influence mechanism.  }
\label{fig3}
\end{figure}

A) The citation attractiveness of an article is an outward manifestation of its novelty, importance and readability\cite{lq10,lq11,lq12}. If an article is more attractive, more valuable, it will receive more citations. In Ref\cite{lq2}, Eom et al give the node (article) a value to represent its citation attractiveness. In this paper, the attractiveness of a node is expressed as a geometric area with the node at center. Since there are highly cited articles each year, the sizes of the influence zones of nodes in the model follow the same power-law distribution.

B) The ``burst'' phenomenon shows that the attractiveness of an article wears off  eventually\cite{lq12,wang1,hajra1,hajra2}. Therefore, the influence zones of nodes in the model shrink with the passage of time.

Combining A and B, an illustration of the model is shown in Figure~\ref{fig3}, and the generating process  is listed as follows.

\begin{enumerate}[Step 1]
  \item  Generate a new circle $C_{t}$ with radius $R(t)=N(t)/(2\pi\delta)$ ($\delta\in\mathbb{R}^{+}$) centered at point $(0,0,t)$ at each time $t=1,2,...T\in\mathbb{Z}^{+}$, sprinkle $N(t)=mt ~(m\in \mathbb{Z}^{+})$ nodes (articles) on it randomly and uniformly, and fix nodes with their coordinates, e.g. node $i$ with $(\theta_{i},t_{i})$.\\
  \item For each node with coordinate $(\theta,t)$, the influence zone of the node is defined as an interval of angular coordinate with center $\theta$  and arc-length   $D=\beta(t)/s(\theta,t)^{\alpha}(t_{c}-t)^{\gamma}$, where $\beta(t)=\beta/t  ~(\beta \in \mathbb{R})$, $\alpha, \gamma\in(0,1)$,  $t_{c}$ is the current time, 
       and $s(\theta,t)$  is  an integer randomly selected from $[1,\eta]$ ($\eta$ is an integer), which corresponds to the node one to one .\\
  \item For node $i$ and node $j$, if the distance of angular coordinates   $\Delta(\theta_{i},\theta_{j})=\pi-|\pi-|\theta_{i}-\theta_{j}||<|D_{i}|$ and $t_{j}>t_{i}$,  a directed edge is drawn from $j$ to $i$ under a probability $p\in[0,1]$.
\end{enumerate}

  Aside from cognition A and B, there are some intuitive explanations for the formula of zonal sizes. Firstly, if the contents of two highly cited articles are concerned with the same hot topics, the cumulative advantage generally would make the older one get more citations. So $\beta(.)$ is set to be inversely proportional to $t$. Secondly, it is easy to understand that different articles have different citation attractiveness, so $s(.,.)$ is adopted into the formula. At last, aging of the citation attractiveness of articles occurs so that the probability of an article receiving citations decreases with its ages, Therefore, the formula is a decreasing function of the node's age $t-t_{c}$, and the parameter $\gamma$ is used to tune the rate at which the formula decreases with $t-t_{c}$.
In addition, the exponents of in-degree distributions of many networks generated by some existing models are fixed\cite{bara1}, which is inconsistent with  the fact that the exponents vary from data to data. So  the parameter $\alpha$ is employed to tune the exponent of the modeled in-degree distributions in the formula of zonal sizes.

Copying citations is considered as a reasonable interpretation for why citation networks have  a non-zero global clustering coefficient\cite{lq13}. The global clustering coefficients in the empirical data are often lower than those predicted by the complete copying, which can be interpreted as a consequence of the aging of the citation attractiveness of an article (older articles hardly get citations). So incomplete copying is more realistic\cite{lq13,lq14,lq15}. In this paper, the model adopts the parameter $p$ to avoid complete copying and simultaneously considers aging of articles (expressed by the sizes of the influence zones of articles which shrink with the passage of time).

\begin{figure}
\centering
 \includegraphics[height=3.2in,width=5.5in,angle=0]{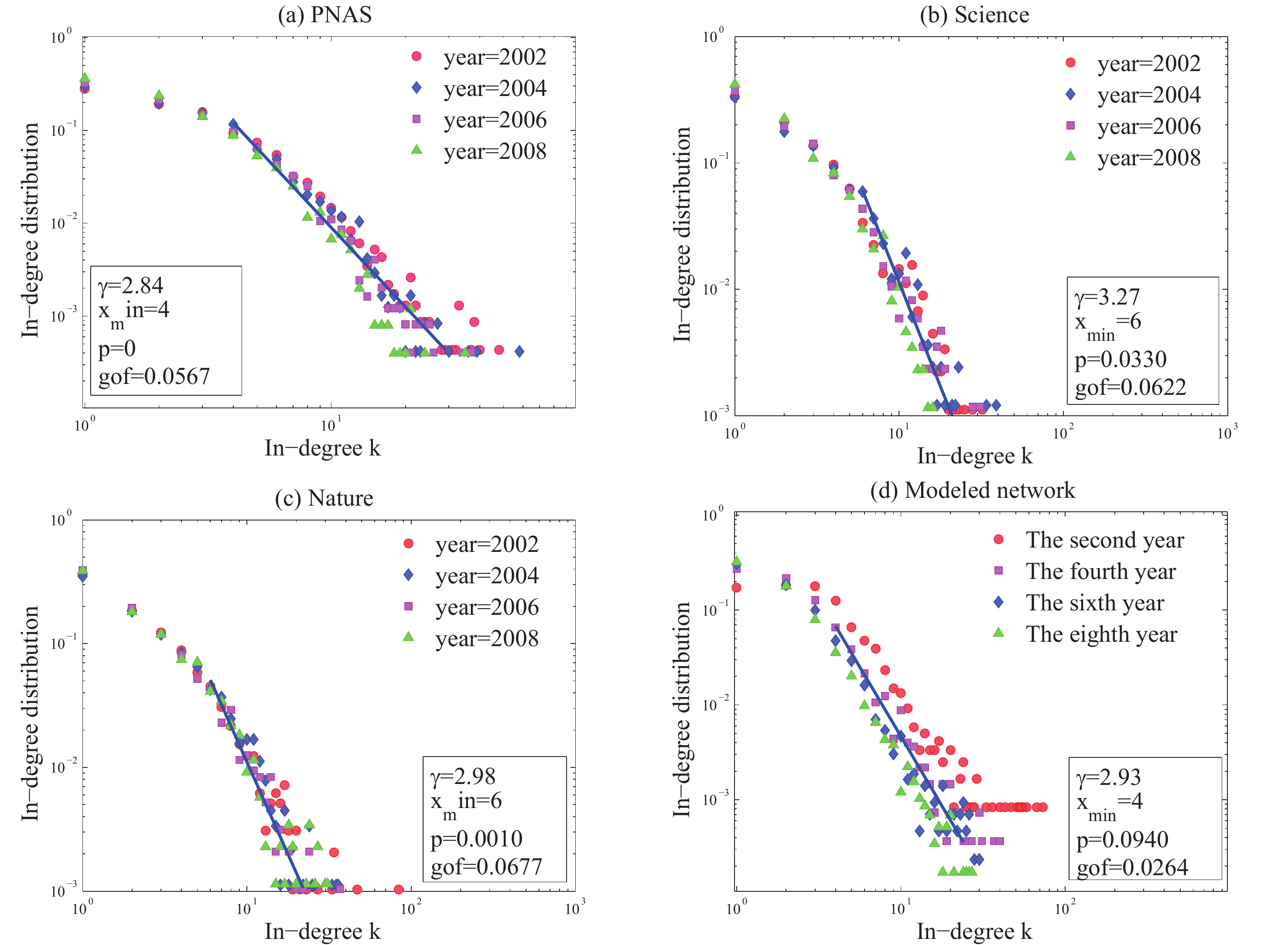}
\caption{\small\textbf{The in-degree distributions of articles published  each year.} Panels show the in-degree distributions of $k-$degree articles of the networks published in four years in Table~\ref{table1}, respectively. The fitting functions are the power-law distribution $p(k)=ak^{-\beta}$ for the tails of the in-degree distributions of articles published in 2004 in the empirical data in Panels (a,b,c) and in the fourth year in the modeled network in  Panel(d) (which is fitted by the method in Ref~\cite{lq16}).}
\label{fig2}       
\end{figure}
 A modeled network (Table~\ref{table1}) is generated  to test  validity of the model to reproduce the properties of PNAS (the fitting function of the annual number of articles in which is $N(t)=123.5t+2719$).
To make the scale consistent with the empirical data, we set $m=12$ and $T=100$.
Moreover, $\varepsilon=8$ time units is regarded as one year so that the time span of the modeled network is nearly equal to 12 ``years'',
which is approximate to that of the empirical data.
 Furthermore, we set other parameters  to make the modeled average degree similar to that of the  empirical data.

\begin{figure}
\centering
 \includegraphics[height=3.2in,width=5.5in,angle=0]{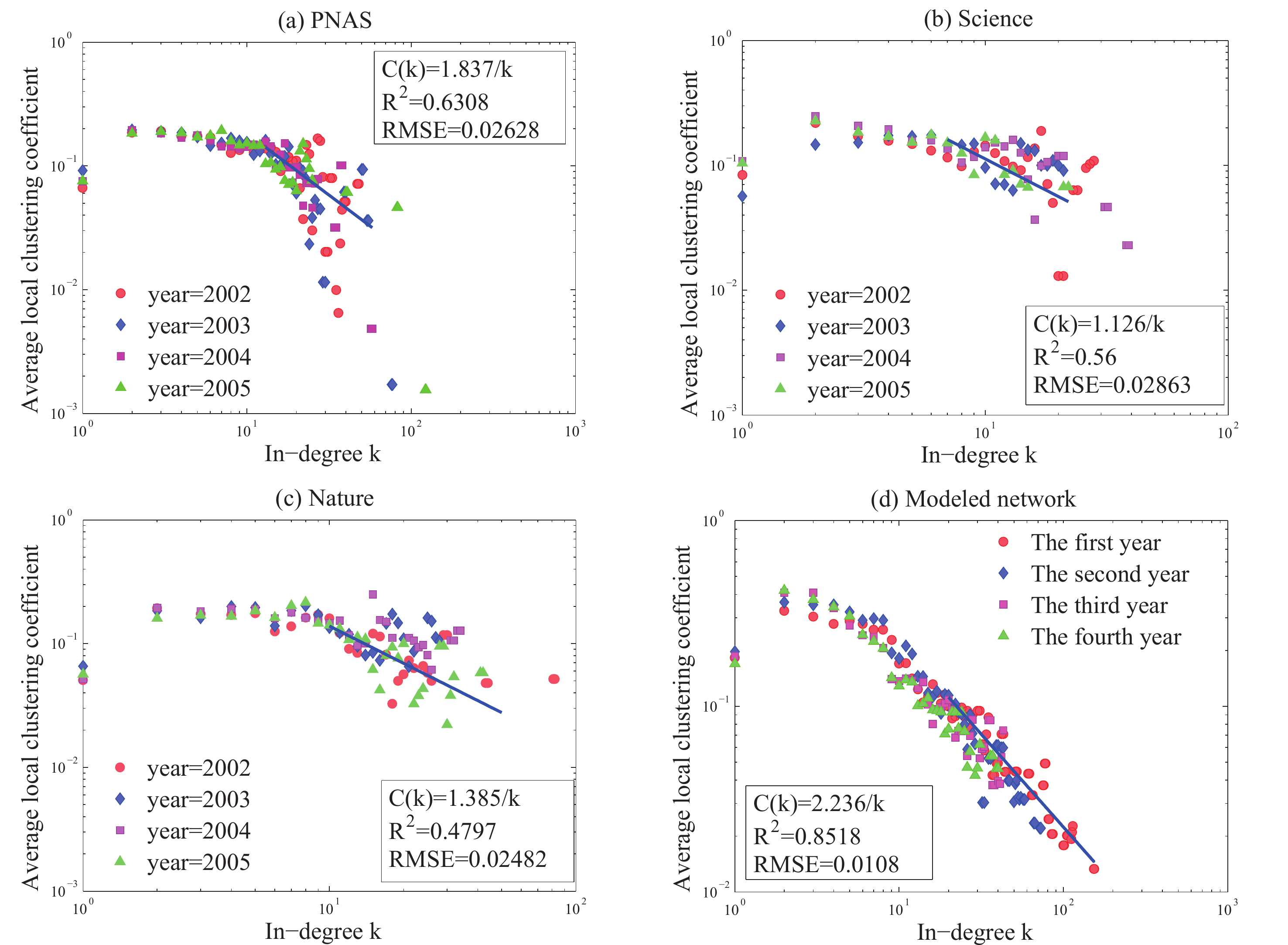}
\caption{\small\textbf{Local clustering coefficients of articles published in four years as functions of in-degree, compared with the theoretical prediction of Eq~\ref{eq14}.}  Panels (a,b,c) show those in  the empirical data, and Panel (d) in the modeled network in Table~\ref{table1}. }
\label{fig5}       
\end{figure}

 Simulations (Figure~\ref{fig2}d) show that the in-degree distributions of articles published each year in the modeled network have similar scale-free tails, which result from the similar inhomogeneous citation attractiveness of articles published each year in the modeled network. Moreover, the``burst'' phenomenon exists in this network (Figure~\ref{fig_burst}d), as a result of aging of the citation attractiveness. In addition, the local clustering coefficient of the modeled network is time-periodic---the tails of the average local clustering coefficients of nodes with in-degree $k$ published each year are roughly proportional to $1/k$ (Figure~\ref{fig5}d).

\section*{Time-periodicity of the in-degree distribution}
The validity of the model (Figure~\ref{fig2}d) in reproducing the time-periodicity of the in-degree distribution in the empirical data convincingly indicates that the existence of the ``burst'' phenomenon and the emergence of hot topics are behind it.
The emergence of hot topics each year results in the similarity of the in-degree distributions in the empirical data. The existence of the ``burst'' phenomenon  increases articles finite, which ensures the scale-free tails of the in-degree distributions.
Accordingly, a further explanation is presented to show why the tails of these distributions are scale-free by analysing and simulating the evolutionary mechanism of citation networks.

The event that whether one article cites another can be regarded as a ``yes/no'' experiment.
So the number of citations an article gets is the number of successes in a sequence of $n$ experiments, where $n$ is the number of articles which are likely to cite this article.

The probability $p$ of ``yes'' is approximated through  its expected value $\hat{p}$, and  those ``yes/no'' experiments are supposed to be independent. Subsequently, the number of citations articles get follows a binomial distribution $B(n,\hat{p})$. When $n$ is large and $\hat{p}$ is small, $B(n,\hat{p})$ can be approximated by a Poisson distribution with mean $n\hat{p}$ (Poisson limit theorem).

Since the process of sprinkling nodes in our model follows a Poisson point process, the number $n$ of nodes located in one influence zone is a random variable drawn from a Poisson distribution with an expected value in proportion to the zonal size, which is different from that of others'.
In addition, the ``yes/no'' experiments could be affected by previous occurrences. For example, articles tend to cite the articles written by distinguished authors.

 Therefore, we should make an approximation. As for articles with large influence zones, the numbers of their potential citers are large enough to suppose that the ``yes/no'' experiments are independent. And the numbers of the citations of these articles could be considered as random variables drawn from a range of Poisson distributions with sufficiently large means.
 A ``fat'' tail appears when  those Poisson distributions are averaged (Eq~(\ref{eq3}) in Appendix).

In the empirical data, since the highly cited articles are generally considered to be of great citation attractiveness, the behaviors that highly cited articles get citations can be regarded as mutually independent. As a result, the tails of the in-degree distributions of articles published each year are scale-free.
It shows that the emergence of scale-free tails partly stems from many Poisson processes, consequently from many ``yes/no'' experiments.

\section*{Time-periodicity of the local clustering coefficient }
The local clustering coefficient (LCC) is tantamount to the probability that two vertices, both neighbors of the third vertex,  will be the neighbors of one another. In the empirical data, the articles published each year exhibit similar feature that the large in-degree articles have low LCCs while the small in-degree ones have high LCCs.  Such time-periodicity of  LCC can be interpreted as a consequence of the time-periodicity of the in-degree distribution. There are both small in-degree articles and large in-degree articles  each year. The small in-degree articles tend to quote the discussion of some highly cited articles in their research fields to enhance the reliability of their points of view. The later articles that cite those small in-degree articles also tend to cite these highly cited articles. So two articles, both neighbors of a small in-degree article, are very likely to be the neighbors of one another.
 On the contrary, the large in-degree articles attract considerate attention of the small in-degree articles, and these large in-degree articles  may not cite each other even if they cite a common article. So the LCCs of the large in-degree articles are low.

Furthermore, considering the average LCC $C(k)$ of nodes  with in-degree $k$ published each year, the tails of $C(k)$ is approximately proportional to $1/k$ (Figure~\ref{fig5}a,\ref{fig5}b,\ref{fig5}c). To show how the model generates a similar tail (Figure~\ref{fig5}d), the formula of the tail of $C(k)$ of the modeled network is derived (Eq~(\ref{eq14}) in Appendix).

\section*{Conclusion}
We have looked at the citation dynamics of  some real networks.  The fact that the in-degree distributions of articles published each year in these networks present similar scale-free tails, leads to the conclusion that their in-degree distributions are time-periodic. The emergence of hot topics and the existence of the ``burst'' phenomenon are cogent reasons behind the generation of time-periodicity of the in-degree distribution. With these two  reasons  synthetically considered, a geometric model based on our previous study is developed.
    The  model boasts two new features: 1) the sizes of the influence zones of nodes follow the same power-law distribution; 2) the sizes of the influence zones of nodes decrease with the passage of time.
   The model manages to  reproduce the time-periodicity  of the in-degree distribution and that of the local clustering coefficient of the empirical data, and accounts for the presence of citation burst as well. Moreover, a sound explanation is presented for the emergence of the scale-free tails of the in-degree distributions of articles published each year by regarding the citation behavior as a ``yes/no'' experiment.  However, some defects of this model have yet to be overcome in future work, including how to identify the coordinates of articles, when the real citation networks are mapped onto the geometric space, and how to design a more realistic strategy for articles to get citations rather than the random and uniform selection strategy employed in the present context.
\appendix
\section{}
The edges in this model are linked according to Step 3.
Then the expected in-degree of node $i$ with coordinate $(\theta_{i},t_{i})$ is
\begin{equation}\label{eq1}
k^{-}(\theta_{i},t_{i})=\sum_{t_{c}=t_{i}+1}^{T}\frac{m\beta pt_{c}}{2\pi s(\theta_{i},t_{i})^{\alpha}(t_{c}-t_{i})^{\gamma}}\approx\frac{m\beta p(T-t_{i})^{2-\gamma}}{2\pi s(\theta_{i},t_{i})^{\alpha}(2-\gamma)t_{i}}+\frac{m\beta p(T-t_{i})^{1-\gamma}}{2\pi(1-\gamma) s(\theta_{i},t_{i})^{\alpha}}.
\end{equation}

 The process of sprinkling nodes in the suppositional year follows Poisson process,  so the actual in-degree of the nodes born in this year are not exactly equal to the expected degree. As Ref~\cite{lq5,liu11} said, we have to average the Poisson distribution,
\begin{align}
p(k^{-}(\theta_{i},t_{i})=k)=\frac{1}{k!}(k^{-}(\theta_{i},t_{i}))^{k}e^{-k^{-}(\theta_{i},t_{i})},
\end{align}
which is the probability that the in-degree of node $i$ is $k$, with the temporal  density $\rho(t_{i})\approx2t_{i}/\delta(2t+\delta)$ in a suppositional year.
Since a suppositional year contains $\varepsilon$ units, the in-degree distribution of the nodes born in a ``year'' from $t$ to $t+\varepsilon$ is
\begin{align}\label{eq3}
p(k^{-}=k)&=\int_{t}^{t+\varepsilon}\rho(t_{i})\frac{1}{\eta}\int_{1}^{\eta}p(k^{-}(\theta_{i},t_{i})=k)dsdt_{i}\nonumber\\
          &=\int_{t}^{t+\varepsilon}\rho(t_{i})\frac{1}{\eta}\int_{1}^{\eta}(\frac{a(t_{i})}{s^{\alpha}})^{k}e^{-\frac{a(t_{i})}{s^{\alpha}}}dsdt_{i}\nonumber\\
          &=\int_{t}^{t+\varepsilon}\rho(t_{i})\frac{a(t_{i})^{\frac{1}{\alpha}}}{\alpha\eta}\int_{\frac{a(t_{i})}{\eta^{\alpha}}}^{a(t_{i})}\frac{1}{k!}\tau^{k-\frac{\alpha+1}{\alpha}}e^{-\tau}d\tau dt_{i}\nonumber\\
          &\approx\int_{t}^{t+\varepsilon}\rho(t_{i})\frac{a(t_{i})^{\frac{1}{\alpha}}}{\alpha\eta}\frac{\Gamma(k-\frac{1}{\alpha})}{\Gamma(k+1)}\int_{\frac{a(t_{i})}{\eta^{\alpha}}}^{a(t_{i})}
          \frac{e^{-\frac{(\tau-k+\frac{\alpha+1}{\alpha})^{2}}{2(k-\frac{\alpha+1}{\alpha})}}}{\sqrt{2\pi(k-\frac{\alpha+1}{\alpha})}}d\tau dt_{i}\nonumber\\
           &\approx\int_{t}^{t+\varepsilon}\rho(t_{i})\frac{a(t_{i})^{\frac{1}{\alpha}}}{\alpha\eta}k^{-(1+\frac{1}{\alpha})}\int_{\frac{a(t_{i})}{\eta^{\alpha}}}^{a(t_{i})}
          \frac{e^{-\frac{(\tau-k+\frac{\alpha+1}{\alpha})^{2}}{2(k-\frac{\alpha+1}{\alpha})}}}{\sqrt{2\pi(k-\frac{\alpha+1}{\alpha})}}d\tau dt_{i},
\end{align}
where $a(t_{i})=m\beta p(T-t_{i})^{2-\gamma}/2\pi(2-\gamma)t_{i}+m\beta p(T-t_{i})^{1-\gamma}/2\pi(1-\gamma)$, $\tau=a(t_{i})/s^{\alpha}$, and $\Gamma(.)$ is the gamma function. Here, we have used the Laplace approximation\cite{liu12} and the Stirling's approximation\cite{liu13} $k!\approx\sqrt{(2\pi k)}(k/e)^{k}$ in the last two steps. Moreover, it can be proved that the integral term of $\tau$ is approximately independent of $k$. The process is as follows:
\begin{align}
\frac{d}{dk}\int_{a(t_{i})}^{\frac{a(t_{i})}{\eta^{\alpha}}}\frac{e^{-\frac{(\tau-k+1+\frac{1}{\alpha})^{2}}{2(k-1-\frac{1}{\alpha})}}}{\sqrt{2\pi(k-1-\frac{1}{\alpha})}}~d\tau
=\frac{e^{-\frac{(\tau-k+1+\frac{1}{\alpha})^{2}}{2(k-1-\frac{1}{\alpha})}}}{\sqrt{2\pi (k-1-\frac{1}{\alpha})}}(1+\frac{\tau}{k-1-\frac{1}{\alpha}})|^{a(t_{i})}_{\frac{a(t_{i})}{\eta^{\alpha}}}\approx 0,
\end{align}
when the in-degree $k$ or $a(t_{i})$ is big enough, the integration is approximately equal to a constant.
Therefore, the in-degree distributions of the nodes born at each suppositional year have fat tails with exponent $1+1/\alpha$.
\section{}
Suppose $i$  is a highly cited article. Article $j$ and article $l$ are the new published articles which are the neighbors of article $i$. If $j$ has coordinate $(\theta_{j},t_{j})$, we make a reasonable assumption that the overlap of the influence zones of $i$ and $j$ in circle $C_{t_{c}}$ ($t_{c}$ is the current time) is approximate $\beta(t_{j})/s(\theta_{j},t_{j})^{\alpha}(t_{c}-t_{j})^{\gamma}$ because of the small $s(\theta_{i},t_{i})$ and large $s(\theta_{j},t_{j})$. Particularly, if the connection probability $p=1$, the probability that article $l$ is the common neighbor of article $i$ and article $j$ is approximately equal to $\beta(t_{j})s(\theta_{i},t_{i})^{\alpha}(t_{c}-t_{i})^{\gamma}/\beta(t_{i})s(\theta_{j},t_{j})^{\alpha}(t_{c}-t_{j})^{\gamma}$. So for the general connection probability $p$, the conditional probability $p(l\mapsto j|l\mapsto i, j\mapsto i)=p\beta(t_{j})s(\theta_{i},t_{i})^{\alpha}(t_{c}-t_{i})^{\gamma}/\beta(t_{i})s(\theta_{j},t_{j})^{\alpha}(t_{c}-t_{j})^{\gamma}$. Summing over possible values of $t_{j}$, we find
\begin{align}\label{eq13}
C(\theta_{i},t_{i})\approx\frac{\int_{t_{i}}^{T}p(l\mapsto j|l\mapsto i, j\mapsto i)\sigma p~dt_{j}}{\int_{t_{i}}^{T}p\sigma ~dt_{j}}
\approx \frac{C(t_{j})s(\theta_{i},t_{i})}{\bar{s_{j}}}
\end{align}
where $\sigma=m\beta(t_{i})t_{j}/2\pi s(\theta_{i},t_{i})^{\alpha}(t_{j}-t_{i})^{\gamma}$  denotes the number of the articles in the influence zone of article $i$ at time $t_{j}$, $C(t_{i})$ is a constant depending on $t_{i}$ and $\bar{s_{j}}$ is the average of $s(\theta_{j},t_{j})$ of all the possible node $j$.

Since article $i$ is a highly cited article, the articles citing $i$ dominate the neighbors of $i$ and the effect of those articles cited by $i$ could be ignored. The expected in-degree of the highly cited article $i$ is  Eq~(\ref{eq1}). Substituting it into Eq~(\ref{eq13}), we get
\begin{align}\label{eq14}
C(k^{-}(\theta_{i},t_{i})=k)\propto\frac{1}{k}
\end{align}
which is inversely proportional to the in-degree $k$ of article $i$.

\end{document}